\documentclass[letter,traditabstract]{aa} 
\usepackage{times,amssymb,amsmath} 
\usepackage{epsfig} 
\usepackage{lscape}

\begin{document} 

\title{Lyman-$\alpha$ emitters as tracers of the transitioning Universe
}

\author{K.K. Nilsson\inst{1,2}
        \and P. M\o ller\inst{2}
}

\institute{
   ST-ECF, Karl-Schwarzschild-Stra\ss e 2, 85748, Garching bei M\"unchen, Germany\\	
\and
     European Southern Observatory, Karl-Schwarzschild-Stra\ss e 2, 85748
   Garching bei M\"unchen, Germany\\
}
\offprints{knilsson@eso.org}
\date{Received date / Accepted date}
\titlerunning{Ly$\alpha$ emitters as tracers of the transitioning Universe}

\abstract
{Of the many ways of detecting high redshift galaxies, the selection of objects due to their redshifted Ly$\alpha$ emission has become one of the most successful. But what types of galaxies are selected in this way? Until recently, Ly$\alpha$ emitters were understood to be small star-forming galaxies, possible building-blocks of larger galaxies. But with increased number of observations of Ly$\alpha$ emitters at lower redshifts, a new picture emerges. Ly$\alpha$ emitters display strong evolution in their properties from higher to lower redshift. It has previously been shown that the fraction of ultra-luminous infrared galaxies (ULIRGs) among the Ly$\alpha$ emitters increases dramatically between redshift three and two. Here, the fraction of AGN among the LAEs is shown to follow a similar evolutionary path. We argue that Ly$\alpha$ emitters are not a homogeneous class of objects, and that the objects selected with this method reflect the general star forming and active galaxy populations at that redshift. Ly$\alpha$ emitters should hence be excellent tracers of galaxy evolution in future simulations and modeling.
}


\maketitle

\section{Introduction}
Several decades of studies of high redshift galaxies have shown that the star formation rate density of the Universe peaked around redshift $z \sim 2$ (e.g. Hogg et al.~1998, Hopkins 2004, Hopkins \& Beacom 2006). At the peak of the star formation history, nearly ten times more stars were formed than in our local Universe, whereas at higher redshifts, the star formation density was equally low as it is now. Similarly, a trend in the volume density of AGN has been found with a peak at $z = 1.5 - 2$ (e.g. Miyaji et al.~2000, Wolf et al.~2003, Bongiorno et al.~2007). The coincidence that the two density functions, for star formation and numbers of AGN, peak at similar redshifts has been proposed as being due to both of these properties being linked to the hierarchical build-up of galaxies and mergers of dark matter haloes (e.g. Kauffmann \& Haehnelt~2000, Bower et al.~2006).

As for the high redshift Universe, one of the strongest emission lines observable is the Lyman-$\alpha$ (Ly$\alpha$) line. By now, several hundreds of Ly$\alpha$ emitters (LAEs) have been detected through narrow-band imaging at $z = 0.3 - 7.7$ (e.g. M{\o}ller \& Warren 1993, Fynbo et al.~2003, Gronwall et al.~2007, Venemans et al.~2007, Nilsson et al.~2007, Finkelstein et al.~2007, Ouchi et al.~2008, Grove et al.~2009, Hibon et al.~2010). Ly$\alpha$ emission may be generated by three main mechanisms; the ionising flux of O and B stars, indicative of star formation, the ionising flux of an energetic UV source, e.g. an active galactic nucleus (AGN), or due to infall of gas on a massive dark matter halo (c.f. Dijkstra et al.~2006a,b, Nilsson et al.~2006). The volume density of sources where the Ly$\alpha$ emission is dominated by the latter is expected to be very low compared to those where the Ly$\alpha$ emission comes from star formation or AGN sources, hence, the volume density of Ly$\alpha$ emitting objects found in the Universe is expected to follow the general evolutionary occurrences of the star formation history, and the AGN history, with redshift.  

In this \emph{Letter} we discuss the fractions of ULIRGs and AGN among Ly$\alpha$ emitters at different redshifts. In a previous publication, the ULIRG fraction among the LAEs was shown to exhibit a sharp transition from very few to a larger sub-sample at a redshift around 2.5 (Nilsson \& M{\o}ller 2009). Here, the fraction of AGN among LAEs is shown to follow a very similar relation, indicating that the underlying galaxy population is transitioning rapidly from $z > 3$ to $z \sim 2$, (see also a similar result in Bongiovanni et al.~2010). We here ask the question how these results relate to the general galaxy evolution in the Universe.

\vskip 5mm
Throughout this paper, we assume a cosmology with $H_0=72$
km s$^{-1}$ Mpc$^{-1}$, $\Omega _{\rm m}=0.3$ and
$\Omega _\Lambda=0.7$.

\section{The AGN fraction of LAEs}\label{sec:fraction}
To determine AGN fractions among LAEs at different redshifts, we started with the data-set of Nilsson et al.~(2009, 2011). This sample of LAEs was found by narrow-band imaging with the ESO2.2m/MPG telescope on La Silla, Chile, using the Wide-Field Imager (WFI). A central section of the COSMOS field was searched for Ly$\alpha$ emitters with $z = 2.206 - 2.312$. Details of the data reduction and candidate selection can be found in Nilsson et al.~(2009), where in total 187 LAE candidates were found. Follow-up spectroscopy of 152 candidates was performed with the VLT/VIMOS instrument early 2010. The results of this campaign will be published in a forthcoming publication, but we will here exclude those candidates which were not confirmed in the spectroscopy. This brings the total sample of LAEs to 171. AGN were selected among the LAEs by use of X-ray data from XMM (Hasinger et al.~2007), \emph{Chandra} (Elvis et al.~2009) and the VLA (Schinnerer et al.~2007). For details, see Nilsson et al.~(2009, 2011). In total, 24 LAEs are selected as AGN. 

\begin{figure}[t]
\begin{center}
\epsfig{file=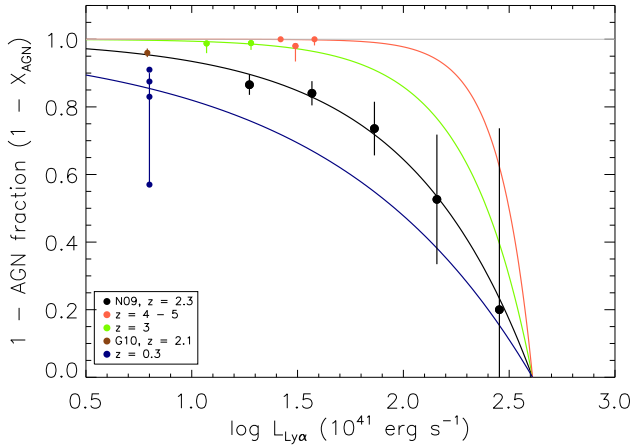,width=9.2cm}
\caption{ Cumulative fraction of non-AGN LAEs as a function of Ly$\alpha$ luminosity. Black points are from the Nilsson et al.~(2009) sample. Coloured points are from publications at redshifts $z=0.3$ (Atek et al. 2009, Cowie et al. 2010, Finkelstein et al. 2009c, Scarlata et al. 2009), $z=2.1$ (Guaita et al.~2010), $z=3$ (Gronwall et al. 2007, Ouchi et al. 2008) and at $z \sim 4-5$ (Wang et al. 2004, Ouchi et al. 2008). Lines are best fit models (see text for details). }
\label{fig:agnfracfunc}
\end{center}
\end{figure}

It was shown already in Nilsson et al.~(2009) that the AGN fraction among these lower redshift galaxies is larger than in samples of LAEs at higher redshifts. With the further \emph{Chandra} detections presented in Nilsson et al.~(2011), this surplus of AGN is even larger. However, since comparing AGN fractions from different surveys has a bias due to the Ly$\alpha$ luminosity reached, a more careful investigation requires an analysis independent of this bias. In Fig.~\ref{fig:agnfracfunc} the AGN fraction is thus presented as a function of Ly$\alpha$ luminosity. The black points are from the Nilsson et al.~(2009) sample.
It is clear that above some flux limit, all Ly$\alpha$ emitting objects are quasars. Similarly, the function tends towards one at fainter fluxes, where the non-AGN LAEs will outnumber the AGN LAEs. 

In Fig.~\ref{fig:agnfracfunc} the results of several other narrow-band surveys for LAEs are also shown. As we have no access to the exact Ly$\alpha$ fluxes of the AGN in most samples, they are shown as single, cumulative points at the Ly$\alpha$ flux limit of each survey. The results include AGN fractions from redshifts $z = 2.1$ (Guaita et al.~2010), $z\sim 3$ (Gronwall et al.~2007, Ouchi et al.~2008), $z = 4 - 5$ (Wang et al.~2004, Ouchi et al.~2008), and at low redshift, $z = 0.3$ (Atek et al.~2009, Cowie et al.~2010, Finkelstein et al.~2009c, Scarlata et al.~2009).

The data-points at $z \sim 2.2$ (the data from the $z=2.3$ sample, and from Guaita et al.~2010) were found to be well fitted with an exponential AGN function of the form:
\begin{equation}  
X_{AGN}(\log L_{Ly\alpha}) = \exp \left( \frac{\log ( L_{Ly\alpha} ) - \log ( L_{*,AGN} ) }{ X_{AGN,0} } \right)
\end{equation}
In this equation, the two free parameters are the $L_{*,AGN}$ which defines the Ly$\alpha$ luminosity where the AGN fraction goes to one, or the other way around that the non-AGN LAE fraction among the narrow-band selected sample goes to zero, and the normalisation constant $X_{AGN,0}$ that determines how quickly the AGN function goes to zero. A further criteria was that the AGN fraction is unity if $\log(L_{Ly\alpha} ) > \log ( L_{*,AGN})$, i.e. that all Ly$\alpha$ emitting objects above a certain Ly$\alpha$ luminosity are AGN-powered. The fits to the $z \sim 2.2$ sample is shown in Fig.~\ref{fig:agnfracfunc}, with best fit parameters $\log ( L_{*,AGN} ) = 43.61^{+0.61}_{-0.27}$~erg~s$^{-1}$ and $X_{AGN,0} = 0.59^{+0.27}_{-0.14}$. For the other redshifts, the $L_{*,AGN}$ was kept constant and only the normalisation constant was determined using all data-points at a particular redshift. This is based on the two assumptions that \emph{i)} the functional form of the AGN fraction is similar at all redshifts, and \emph{ii)} that the Ly$\alpha$ luminosity above which all emitters are AGN does not change with redshift. These fits are also shown in the Figure. A clear evolution from higher to lower redshift appears to be present. 

\section{Evolution in the AGN fraction with time}
\begin{figure}[t]
\begin{center}
\epsfig{file=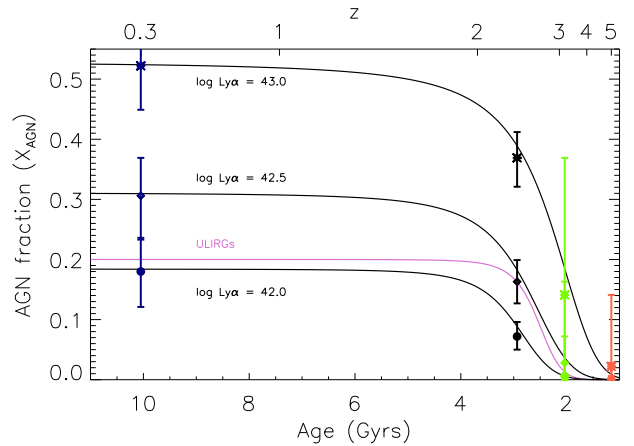,width=9.2cm}
\caption{ Evolution in AGN fraction above a given Ly$\alpha$ luminosity as a function of age of the Universe. Colour scheme is identical to Fig.~\ref{fig:agnfracfunc} and refers to redshift. Different symbols mark the AGN fractions above three representative Ly$\alpha$ luminosities. The black lines are fits to the data, see text for explanation. A transition redshift at $z \sim 2.5$ (age~$\sim 2.5$~Gyrs) are seen in all fits independent of luminosity limit. The purple line shows the redshift transition of the fraction of ULIRGs in LAE samples as presented in Nilsson \& M{\o}ller~(2009).  }
\label{fig:agnevol}
\end{center}
\end{figure}

To illustrate the evolution of the LAE AGN fraction with time, the fraction of AGN LAEs in the sample was extracted from the fits at three different Ly$\alpha$ luminosities,  i.e. the AGN fraction that a given exponential fit in Fig.~\ref{fig:agnfracfunc} gives for a certain Ly$\alpha$ luminosity, and plotted as a function of redshift in Fig.~\ref{fig:agnevol}. 
Worst case uncertainties on these values were found by varying the exponential fits within their $1\sigma$ uncertainties and finding the minimum/maximum AGN fraction expected.
In this Figure, a clear trend towards larger AGN fractions at lower redshifts is seen, and in particular a rather sudden transition from very small fractions at $z \gtrsim 3$ to significant fractions at $z \lesssim 2$. This trend mimics the trend of increasing ULIRG fraction among the LAEs with decreasing redshift, as found in Nilsson \& M{\o}ller (2009; also plotted in Fig.~\ref{fig:agnevol}). The AGN data were fitted, minimising the $\chi^2$ function, with the same transitional equation as in Nilsson \& M{\o}ller (2009):
\begin{equation}\label{eq:trans}
AF(z) = \frac{AF_{0}}{2} (1-\tanh(\theta \,\, (z - z_{tr})))
\end{equation}
In this equation, $AF$ is the AGN fraction, $\theta$ represents
the steepness of the transition and $z_{tr}$ is the transition redshift. The best fit values for the different Ly$\alpha$ luminosity limits can be found in Table~\ref{tab:fits}. The most interesting parameter is the transition redshift. It appears that the brightest AGN transition first, while fainter AGN remain less numerous until later times. This is consistent with a down-sizing scenario, where the larger galaxies/AGN form first (Juneau et al.~2005). 
Comparing the results with those for the LAE ULIRGs shows the transition for the ULIRGs to be sharper than those for the AGN, with a transition redshift nearer to those of the brighter LAE AGN. 
\begin{table}[t]
\begin{center}
\caption{Best fit results for Eq.\ref{eq:trans}. }
\begin{tabular}{@{}lcccccc}
\hline
\hline
$\log$~L$_{\mathrm{Ly}\alpha}$ & $z_{tr}$ & AF$_0$ & $\theta$ \\
\hline
42.0       & 2.18$^{+0.33}_{-0.73}$ & 0.18$^{+0.06}_{-0.06}$ & 1.84$^{+1.15}_{-1.18}$ \\
42.5       & 2.33$^{+0.39}_{-0.52}$ & 0.31$^{+0.08}_{-0.07}$ & 1.31$^{+0.59}_{-0.77}$ \\
43.0       & 2.76$^{+0.52}_{-0.24}$ & 0.53$^{+0.09}_{-0.07}$ & 0.91$^{+0.67}_{-0.51}$ \\
\hline
ULIRGs & 2.52 & 0.2   & 2.28 \\
\hline
\label{tab:fits}
\end{tabular}
\end{center}
\begin{list}{}{}
\item[\textbf{Notes.}] The results at the bottom of the table refer to the best \\ fit parameters for a sample of LAE ULIRGs, presented in Nilsson \\ \& M{\o}ller (2009).
\end{list}
\end{table}

\section{Discussion}\label{sec:results}
\subsection{AGN and LAE volume density evolution}
\begin{figure*}[t]
\begin{center}
\epsfig{file=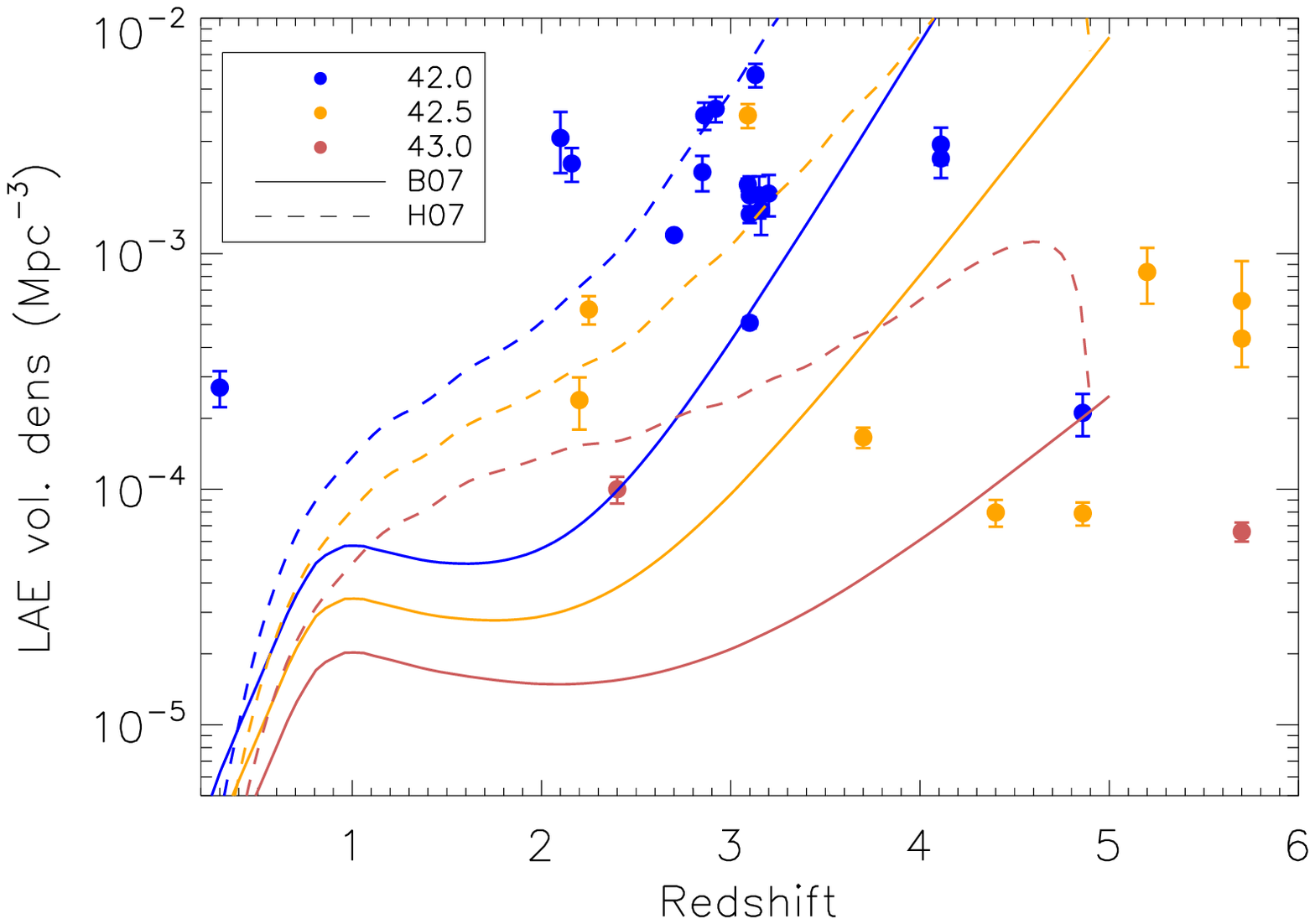,width=9.0cm}\epsfig{file=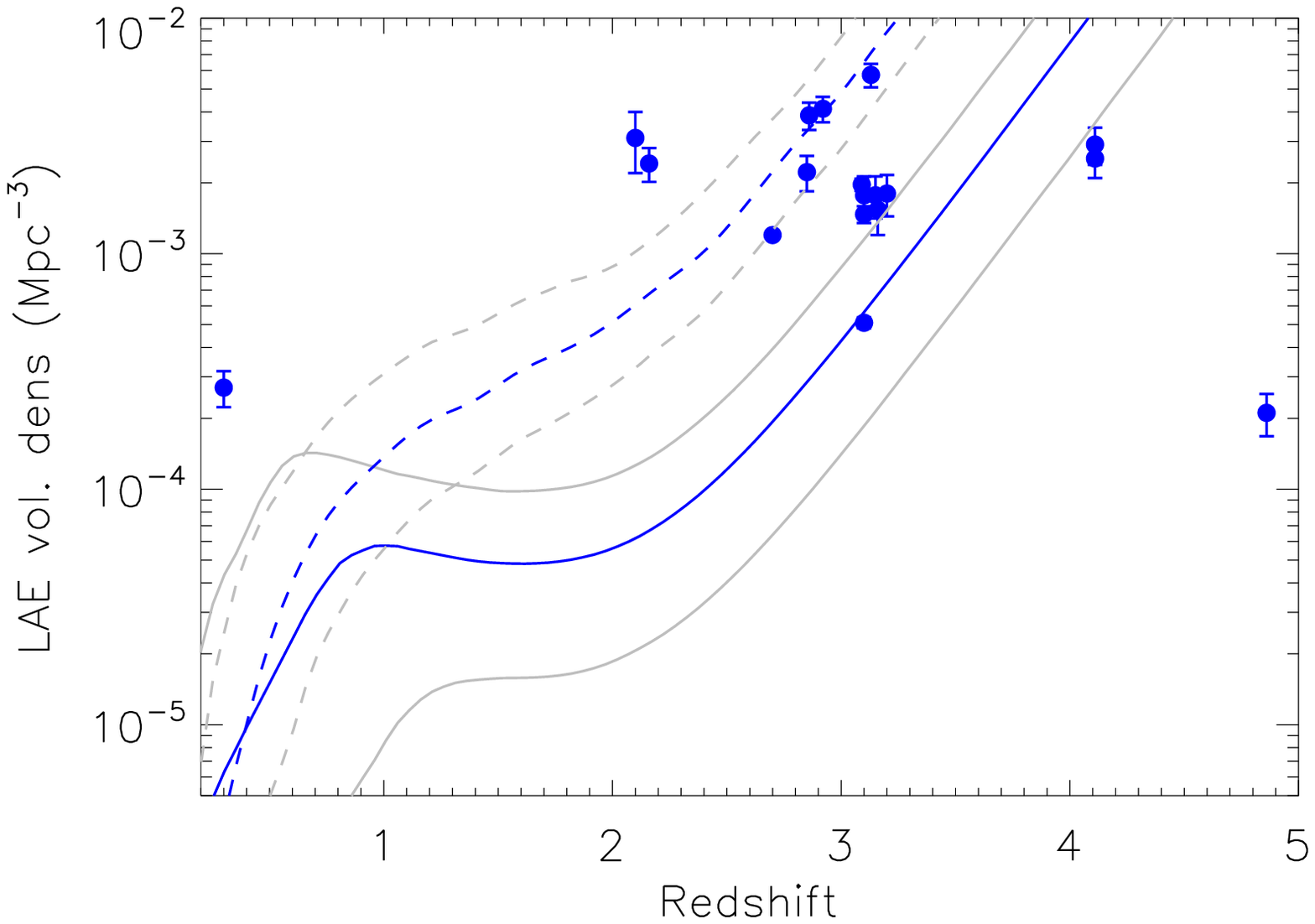,width=9.0cm}
\caption{ LAE volume densities as a function of redshift and limiting Ly$\alpha$ luminosity. The data-points are collected from 18 publications, searching for LAEs, with colour coding referring to the limiting Ly$\alpha$ luminosity in each survey (where [blue, yellow, red] have $\log \,\, \mathrm{Ly}\alpha$~[$< 42.25$, $42.25 - 42.75$, $>42.75$]~erg~s$^{-1}$). Models (see text) come from Bongiorno et al.~(2007; B07) and Hopkins et al.~(2007; H07). To the left, all measurements, and the two models (with M$_B < -22.0$), are shown to the different Ly$\alpha$ luminosity limits. Beyond $z\sim 4.5$ both models and measurements become uncertain. At $z = 2-3$ the measurements agree reasonably well with the expectations. The volume density at $z = 0.3$ is significantly too large, compared to the models. To the right, only models and measurements to the faintest luminosity limits are shown, and instead the effect of varying the luminosity of the AGN is shown. For each model the gray lines show the predictions for M$_B < -20.0$ (above the original line) or M$_B < -24.0$ (below the original line). The volume density at $z = 0.3$ agrees better now, if all AGN at this redshift are very faint in the continuum.  }
\label{fig:laesurfdens}
\end{center}
\end{figure*}

Having determined the AGN and ULIRG fractions among LAEs, one may ask the question how these fractions relate to the general evolution of galaxy properties with redshift. An independent test of the determined AGN fraction is to calculate the LAE volume density, by dividing a known AGN volume density by the LAE AGN fraction. For this test, we use the AGN comoving volume densities as determined observationally by Bongiorno et al.~(2007) and theoretically by Hopkins et al.~(2007), and we compare with comoving LAE volume densities from eighteen different publications (Steidel et al.~2000, Stiavelli et al.~2001, Dawson et al.~2004, Hayashino et al.~2004, Shimasaku et al.~2006, Nilsson et al.~2007, Gronwall et al.~2007, Venemans et al.~2007, Murayama et al.~2007, Prescott et al.~2008, Deharveng et al.~2008, Ouchi et al.~2008, Nilsson et al.~2009, Grove et al.~2009, Shioya et al.~2009, Guaita et al.~2010, Bongiovanni et al.~2010, Yuma et al.~2010). In Fig.~\ref{fig:laesurfdens} these LAE volume densities are shown, together with the predictions.
In the following, all Type 1 AGN are assumed to have strong enough Ly$\alpha$ emission lines to be detected as LAEs. The predictions in the left panel are done for AGN brighter than M$_B < -22$, as the $z = 2.25$ LAE AGN from Nilsson et al.~(2009) are all brighter than this magnitude. The scatter in the data-points is large, probably due to different selection criteria for the different surveys, but the overall trends are clear; the LAE volume density remains relatively flat over the redshift range $z = 2 - 6$, but falls at low redshift. The volume density of LAEs predicted by the AGN fraction, in turn, has a steep increase towards higher redshift. In the range $z = 2 - 3$, the predictions and the measurements agree well, but there is an apparent lack/surplus of LAEs at high/low redshift. One explanation to this trend may be that the LAE AGN have different luminosities at different redshifts. In the right panel of Fig.~\ref{fig:laesurfdens} the effect of varying the AGN luminosity is shown. It becomes clear that the lack of LAEs at very high redshift ($z>4$) can be explained if the AGN are all very luminous, and similarly that the surplus of AGN at $z=0.3$ can be explained if all AGN are faint. The fact that the LAE volume density agrees so well with the predictions is an independent confirmation of the derived AGN fraction evolution, as well as an indication that the LAE AGN follow the general AGN evolutionary trends in the Universe. 

\subsection{Ly$\alpha$ emitters as tracers of galaxy evolution}
The apparent trends displayed by LAEs as a function of redshift are easily explained with a simple, phenomenological model when considering which objects in the Universe emit Ly$\alpha$ photons. At high redshift ($z > 3$), most galaxies have low stellar masses and are in their first star forming episode. AGN and ULIRG number densities are low. Most Ly$\alpha$ photons at this time will thus trace this; young, small star-forming galaxies in their first starburst, with little dust content. Very few AGN or ULIRG LAEs will be found. A Gyr later, at $z \sim 2$, both the star formation rate density and the AGN number density are peaking. Some galaxies are already in their second, or later, starburst. Now the sample of Ly$\alpha$ selected galaxies will reflect this diversity, with more AGN and ULIRG LAEs detected and a larger range of stellar properties (cf.~Pentericci et al.~2009, Nilsson et al.~2011). Whereas there will be some galaxies in the same stage of evolution as at higher redshift, there will also be some galaxies with evolved stellar populations that are experiencing more recent star formation. These galaxies can be more massive and/or more dusty. The study of the redshift evolution of LAEs can thus be seen as an evolution in Ly$\alpha$ emission in the Universe, and could as such be an interesting test for galaxy evolution models including both star formation, dust effects and AGN number densities. Note that in this model, no assumption about the evolution of individual LAEs is made. It is unlikely that LAEs at higher redshift will appear as such at lower redshift. The argument here is rather that LAEs are a random sub-set of all star-forming or active galaxies at a certain time, that happen to be in a Ly$\alpha$ emitting phase. Also, from this data it is not possible to draw any conclusions on the evolution of passive (i.e. non-star-forming or inactive) galaxies.

\section{Conclusion}
Based on the apparent evolution of the AGN and ULIRG fractions of LAEs, fractions that are expected to follow the evolution of star forming and active galaxy populations throughout the Universe, some conclusions can be drawn regarding the general galaxy evolution scenario. From the evolutions of the fractions and volume densities in Fig.~\ref{fig:agnevol} and \ref{fig:laesurfdens} it is seen that at first, in the very young ($z > 4$) Universe, only stars formed, regardless of AGN or ULIRGs. Later, at an age of approximately 3 Gyrs or redshift $\sim 2.5$, a secondary process started, suddenly increasing the number of both AGN and ULIRGs in the galaxy population. After this abrupt increase in dusty and active galaxies, an equilibrium was reached that has lasted to this day. A possible explanation for this very stable equilibrium may be due to feed-back effects controlling the SFR in the galaxies. It is clear that this remains speculation until further data is collected, but proves the importance of studying objects emitting Ly$\alpha$ emission in the Universe.

\begin{acknowledgements}
The authors wish to thank Lucia Guaita for providing information about the AGN in CDFS and Yujin Yang for useful comments on the draft.
\end{acknowledgements}

\end{document}